\documentclass[a4paper,comsoc]{IEEEtran}
\IEEEoverridecommandlockouts

\usepackage{cite}
\usepackage{amsmath,amssymb,amsfonts}
\usepackage{algorithmic}
\usepackage{algorithm}
\usepackage{graphicx}
\usepackage{textcomp}
\usepackage{xcolor}
\usepackage{siunitx}
\usepackage{tikz}
\usepackage{pgfplots}
\usepgfplotslibrary{statistics}
\usetikzlibrary{arrows,shapes}
\usetikzlibrary{decorations.pathmorphing,calc,shapes,shapes.geometric,patterns}
\pgfplotsset{compat=1.8}
\usepackage{caption}
\usepackage{balance}

\tikzset{gmm/.style={mark options={solid},color=TUMBeamerBlue, line width=\lineWidth, mark=square, mark size=\marksize, solid}}
\tikzset{gmmtoep/.style={mark options={solid},color=TUMMediumGray, line width=\lineWidth, mark=triangle, mark size=\marksize, solid}}
\tikzset{lmmse/.style={mark options={solid},color=TUMBeamerOrange, mark = o,  line width=\lineWidth, dashed}}
\tikzset{cnn/.style={mark options={solid},color=black, line width=\lineWidth, mark=x, mark size=\marksize, dotted}}

\usepackage{subcaption}

\definecolor{myblack}{RGB}{70,70,70}
\definecolor{myblue}{RGB}{65,105,225}
\definecolor{mygreen}{RGB}{0,139,139}
\definecolor{myorange}{RGB}{255,150,0}
\definecolor{myred}{RGB}{255,69,0}
\definecolor{mylila}{RGB}{153,50,204}

\newcommand{\lineWidth}{1.0pt}
\newcommand{\marksize}{2.0pt}

\tikzset{cnn/.style={mark options={solid},color=myred, line width=\lineWidth, mark=star, mark size=\marksize, dashdotted}}
\tikzset{gmm/.style={mark options={solid},color=TUMBeamerBlue, line width=\lineWidth, mark=square, mark size=\marksize}}
\tikzset{lmmse/.style={mark options={solid},color=black, mark = o,  line width=\lineWidth, dashed}}

\tikzset{jakes20/.style={mark options={solid},color=TUMBeamerOrange, line width=\lineWidth, mark=triangle, mark size=\marksize, dotted}}
\tikzset{jakes10/.style={mark options={solid},color=TUMMediumGray, line width=\lineWidth, mark=diamond, mark size=\marksize, dotted}}
\tikzset{jakes/.style={mark options={solid},color=mylila, line width=\lineWidth, mark=asterisk, mark size=\marksize, dotted}}
\tikzset{gmmtoep/.style={mark options={solid},color=TUMBeamerGreen, line width=\lineWidth, mark=pentagon, mark size=\marksize}}

\def\BibTeX{{\rm B\kern-.05em{\sc i\kern-.025em b}\kern-.08em
    T\kern-.1667em\lower.7ex\hbox{E}\kern-.125emX}}

\usepackage{mathtools}	
\usepackage{amssymb}
\usepackage{amsthm}

\usepackage[utf8]{inputenc}

\usepackage{ifthen}

\usepackage{microtype}

\usepackage{caption}

\usepackage{bm}

\usepackage[draft,nomargin,marginclue,inline]{fixme}
\makeatletter
\renewcommand*\FXLayoutMarginClue[3]{%
  \marginpar[%
  \raggedleft\@fxuseface{margin}\textcolor{red}{\ignorespaces $ \Rightarrow $}]{%
    \raggedright\@fxuseface{margin}\textcolor{red}{\ignorespaces $ \Leftarrow $}}}
\makeatother	

\usepackage{tumcolor}

\usepackage{pgfplotstable}	

\pgfplotsset{
	discard if/.style 2 args={
        x filter/.append code={
            \edef\tempa{\thisrow{#1}}
            \edef\tempb{#2}
            \ifx\tempa\tempb
                
            \fi
        }
    },
    discard if not/.style 2 args={
        x filter/.append code={
            \edef\tempa{\thisrow{#1}}
            \edef\tempb{#2}
            \ifx\tempa\tempb
            \else
                
            \fi
        }
    }
}

\usepackage{cite}

\usepackage[acronym,shortcuts]{glossaries}
\newacronym{cnn}{CNN}{convolutional neural network}
\newacronym{ula}{ULA}{uniform linear array}

\usepackage{cleveref}

\usepackage{enumitem}

\tikzset{algorithm1/.style={mark options={solid},color=TUMBeamerBlue, line width=\lineWidth, mark=square, dashed}}

\pgfplotscreateplotcyclelist{miko}{%
{color=TUMBeamerRed, dash pattern=on 5pt off 1pt on 1pt off 1pt, line width=1.5pt},%
{color=TUMBeamerOrange, dash pattern=on 5pt off 2pt, line width=1.5pt,
mark=o},%
{color=black, dash pattern=on 5pt off 2pt on 3pt off 2pt, line width=1.5pt},%
{color=TUMDarkerBlue, line width=1.5pt},%
{color=black, dotted, line width=1.5pt},%
{color=TUMBeamerGreen, dotted, line width=1.5pt},%
{color=TUMBeamerYellow, dotted, line width=1.5pt},%
{color=TUMBeamerDarkRed, dotted, line width=1.5pt},%
{color=TUMBeamerLightGreen, dotted, line width=1.5pt},%
{color=TUMIvory, dotted, line width=1.5pt},%
{color=TUMLightBlue, dotted, line width=1.5pt},%
}

\DeclareMathOperator{\diag}{diag}

\DeclareMathOperator{\expec}{E}

\newcommand*{\mc}[1]{\mathcal{#1}}	

\newcommand{\calN}{\mathcal{N}}
\newcommand{\calO}{\mathcal{O}}

\newcommand*{\C}{\mathbb{C}}

\newcommand*{\R}{\mathbb{R}}

\newcommand{\herm}{{\operatorname{H}}}
\newcommand{\tp}{{\operatorname{T}}}

\definecolor{myblue}{RGB}{30, 100, 200}

\newlength{\leftstackrelawd}
\newlength{\leftstackrelbwd}
\def\leftstackrel#1#2{\settowidth{\leftstackrelawd}%
	{${{}^{#1}}$}\settowidth{\leftstackrelbwd}{$#2$}%
	\addtolength{\leftstackrelawd}{-\leftstackrelbwd}%
	\leavevmode\ifthenelse{\lengthtest{\leftstackrelawd>0pt}}%
	{\kern-.5\leftstackrelawd}{}\mathrel{\mathop{#2}\limits^{#1}}}

\newcommand{\mbC}{\bm{C}}

\newcommand{\mbQ}{\bm{Q}}

\newcommand{\mbS}{\bm{S}}

\newcommand{\mbc}{\bm{c}}

\newcommand{\mbh}{\bm{h}}

\newcommand{\mbn}{\bm{n}}

\newcommand{\mbw}{\bm{w}}

\newcommand{\mby}{\bm{y}}

\newcommand{\mbSigma}{{\bm{\Sigma}}}

\newcommand{\mbmu}{{\bm{\mu}}}

\newcommand{\covhi}{\mbC_i}
\newcommand{\covhk}{\mbC_k}
\newcommand{\meanhi}{\mbmu_i}
\newcommand{\meanhk}{\mbmu_k}

\Crefname{figure}{Fig.}{Figs.}

\newacronym{AWGN}{AWGN}{additive white Gaussian noise}
\newacronym{BLMMSE}{BLMMSE}{Bussgang LMMSE}
\newacronym{BS}{BS}{base station}
\newacronym{CDF}{CDF}{cumulative distribution function}
\newacronym{CNN}{CNN}{convolutional neural network}
\newacronym{CSI}{CSI}{channel state information}
\newacronym{CSIT}{CSIT}{channel state information at the transmitter}
\newacronym{DFT}{DFT}{discrete Fourier transform}
\newacronym{DL}{DL}{downlink}
\newacronym{DNN}{DNN}{deep neural network}
\newacronym{DoA}{DoA}{direction of arrival}
\newacronym{EM}{EM}{expectation-maximization}
\newacronym{FDD}{FDD}{frequency division duplex}
\newacronym{GMM}{GMM}{Gaussian mixture model}
\newacronym{LMMSE}{LMMSE}{linear minimum mean squared error}
\newacronym{LOS}{LOS}{line of sight}
\newacronym{LS}{LS}{least squares}
\newacronym{MSE}{MSE}{mean squared error}
\newacronym{MIMO}{MIMO}{multiple-input multiple-output}
\newacronym{MPC}{MPC}{multi-path component}
\newacronym{MT}{MT}{mobile terminal}
\newacronym{NLOS}{NLOS}{non-line of sight}
\newacronym{NN}{NN}{neural network}
\newacronym{O2I}{O2I}{outdoor-to-indoor}
\newacronym{OMP}{OMP}{orthogonal matching pursuit}
\newacronym{PDF}{PDF}{probability density function}
\newacronym{PGD}{PGD}{projected gradient descent}
\newacronym{PSD}{PSD}{power spectral density}
\newacronym{SNR}{SNR}{signal-to-noise ratio}
\newacronym{TDD}{TDD}{time division duplex}
\newacronym{UL}{UL}{uplink}
\newacronym{ULA}{ULA}{uniform linear array}
\newacronym{URA}{URA}{uniform rectangular array}
\newacronym{UMa}{UMa}{urban macrocell}
\newacronym{UMi}{UMi}{urban microcell}
\newacronym{nSE}{nSE}{normalized spectral efficiency}
\newacronym{cCDF}{cCDF}{complementary cumulative distribution function}
\newacronym{CME}{CME}{conditional mean estimator}
\newacronym{OFDM}{OFDM}{orthogonal frequency-division multiplexing}
\newacronym{LTE}{LTE}{Long Term Evolution}
\newacronym{GPS}{GPS}{Global Positioning System}
\newacronym{ce}{CE}{channel estimation}
\newacronym{ml}{ML}{machine learning}
\newacronym{phy}{PHY}{physical layer}
\newacronym{AGV}{AGV}{automated guided vehicle}

\pgfplotsset{compat=1.15}

\begin{document}

\title{Wireless Channel Prediction via\\ Gaussian Mixture Models}

\author{Nurettin Turan\IEEEauthorrefmark{1}, Benedikt Böck\IEEEauthorrefmark{1}, Kai Jie Chan\IEEEauthorrefmark{1}, Benedikt Fesl\IEEEauthorrefmark{1}, Friedrich Burmeister\IEEEauthorrefmark{2}, Michael Joham\IEEEauthorrefmark{1},\\ Gerhard Fettweis\IEEEauthorrefmark{2}, and Wolfgang Utschick\IEEEauthorrefmark{1}\\
\IEEEauthorblockA{\IEEEauthorrefmark{1}TUM School of Computation, Information and Technology, Technische Universität München, Germany\\
\IEEEauthorrefmark{2} Vodafone Chair Mobile Communications Systems, Technische Universität Dresden, Germany\\
Email: nurettin.turan@tum.de}
\thanks{The authors acknowledge the financial support by the Federal Ministry of Education and Research of Germany in the program of “Souverän. Digital. Vernetzt.”. Joint project 6G-life, project identification numbers: 16KISK001K and 16KISK002.}
\thanks{\copyright This work has been submitted to the IEEE for possible publication. Copyright may be transferred without notice, after which this version may no longer be accessible.
}
}

\maketitle

\begin{abstract}
In this work, we utilize a \ac{GMM} to capture the underlying \ac{PDF} of the channel trajectories of moving \acp{MT} within the coverage area of a \ac{BS} in an offline phase.
We propose to leverage the same \ac{GMM} for channel prediction in the online phase.
Our proposed approach does not require \ac{SNR}-specific training and allows for parallelization. 
Numerical simulations for both synthetic and measured channel data demonstrate the effectiveness of our proposed \ac{GMM}-based channel predictor compared to state-of-the-art channel prediction methods.

\end{abstract}


\section{Introduction}

To achieve high data rates in wireless communications systems, the knowledge of \ac{CSI} at the \ac{BS} is a crucial prerequisite.
In scenarios where the \acp{MT} are moving along trajectories, the \ac{CSI} knowledge gets outdated. 
Therefore, accurate \ac{CSI} prediction, which aims to forecast upcoming \ac{CSI} given past observations corrupted by noise, is highly important.

Classical techniques that utilize linear prediction filters were proposed in, e.g., \cite{Zemen, BaBe05, Kay}.
More recently, non-linear techniques, especially \ac{NN}-based solutions, were proposed for channel prediction in, e.g., \cite{JiCuNgDa22, YuNgMa19, TuUt20}.
Both classical and \ac{NN}-based techniques require the velocity knowledge of the moving \ac{MT} and/or are specifically trained for a particular \ac{SNR} level.
In the case of the \ac{NN}-based approaches, in addition to the burden of training many specialized \acp{NN}, a separate network needs to be stored for each velocity and/or \ac{SNR} configuration.

Recently, \acp{GMM} (cf. \cite[Sec.~9.2]{bookBi06}) were utilized to capture the underlying \ac{PDF} of any channel that stems from a particular communications environment and were used for, e.g., channel estimation in \cite{KoFeTuUt21J} and precoding in \cite{TuFeKoJoUt23}.
Motivated by the universal approximation property of \acp{GMM} (see \cite{NgNgChMc20}), we propose to utilize \acp{GMM} for channel prediction in this work to address the aforementioned drawbacks of classical and \ac{NN}-based techniques.

\emph{Contributions:}
We utilize the discrete latent space of GMMs to derive an approximate minimum \ac{MSE}-optimal channel predictor for given noisy observations that is composed of an observation-dependent convex combination of \ac{LMMSE} prediction filters.
A key feature of our proposed GMM predictor is that after fitting a single GMM at the \ac{BS} to a training set of trajectories with a predefined length in the offline phase, the same GMM can be used online for channel prediction with customized observation and prediction intervals.
Moreover, the \ac{GMM} supports channel prediction for any desired \ac{SNR} level and neither requires retraining nor the knowledge of the velocities of the \acp{MT}.
Numerical results for both synthetic and measured channel data demonstrate the superior performance of our proposed method compared to state-of-the-art channel prediction approaches.

\section{System Model and Channel Data}

The channel coefficients along a trajectory of a moving \ac{MT} are denoted by $h[m]$, with $m=0,\dots, M_\mathrm{o}+N_\mathrm{p}-1$, where $M_\mathrm{o}$ is the \textit{observation length}, and $N_\mathrm{p}$ is the \textit{prediction length}.
During the symbol duration $T_\mathrm{S}$, the channel coefficients $h[m]$ are assumed to remain constant.
The goal of this work is to predict any desired channel coefficient out of the prediction interval $\mathcal{I}_{N_\mathrm{p}} = \{M_\mathrm{o},M_\mathrm{o}+1,\dots, M_\mathrm{o}+N_\mathrm{p}-1\}$, given noisy observations of the channel coefficients of the observation interval $\mathcal{I}_{M_\mathrm{o}} = \{0,1,\dots, M_\mathrm{o}-1\}$.
Thus, the \ac{BS} receives
\begin{equation} \label{eq:signal_model}
    \mby = \mbh_{M_\mathrm{o}} + \mbn \in \C^{M_\mathrm{o}}
\end{equation}
where $\mbh_{M_\mathrm{o}} = [h[M_\mathrm{o}-1], h[M_\mathrm{o}-2], \dots, h[1], h[0]]^\tp$ comprises the channel coefficients of the observation interval $\mathcal{I}_{M_\mathrm{o}}$, and $\mbn \sim \mathcal{N}_\C(\mathbf{0}, \mbSigma = \sigma^2 \mathbf{I}_{M_\mathrm{o}})$ denotes \ac{AWGN}.

With $\mbh$, we denote the vector of channel coefficients of the observation interval $\mathcal{I}_{M_\mathrm{o}}$ extended by all of the coefficients of the prediction interval $\mathcal{I}_{N_\mathrm{p}}$, i.e., $\mbh = [{h[M_\mathrm{o}+N_\mathrm{p}-1]},{h[M_\mathrm{o}+N_\mathrm{p}-2]},\dots,{h[M_\mathrm{o}]},\mbh_{M_\mathrm{o}}^\tp]^\tp \in \C^{M_\mathrm{o}+N_\mathrm{p}}$.
We have that
\begin{equation} \label{eq:sel_matrix}
    \mbh_{M_\mathrm{o}} = \mbS^\tp  \mbh
\end{equation}
with the selection matrix $ 
       \mbS = 
       \begin{bmatrix}
        \mathbf{0},
        \mathbf{I}_{M_\mathrm{o}}
        \end{bmatrix}^\tp
\in \{0,1\}^{M_\mathrm{o}+N_\mathrm{p} \times M_\mathrm{o}}$.

With
\begin{equation} \label{eq:H_dataset}
     \mathcal{H} = \{ \mbh_j \}_{j=1}^{J},
\end{equation}
we denote the training data set consisting of $J$ trajectories.
We consider two different data sources in this work, briefly outlined next.

\subsection{Synthetic Channel Data} \label{sec:syn_data}

We utilize the QuaDRiGa channel simulator \cite{QuaDRiGa1, QuaDRiGa2} to generate channel trajectories in an \ac{UMa} scenario.
We consider a single-carrier with a carrier frequency $f_\mathrm{c} =\SI{3.5}{\giga\hertz}$ and the symbol duration is $T_\mathrm{S}=\SI{0.5}{ms}$.
Placed at a height of $\SI{20}{\meter}$, the \ac{BS} covers a $120\SI{}{\degree}$ sector.
The \acp{MT} are outdoors and move along straight trajectories with a constant velocity $v$, uniformly drawn from 3 to 100~km/h, at a height of $\SI{1.5}{\meter}$.
The \ac{BS} and each \ac{MT} are equipped with single antennas.
The distances between the \acp{MT} and the \ac{BS} are randomly drawn between $\SI{35}{\meter}$ and $\SI{500}{\meter}$. Following the description in the QuaDRiGa manual~\cite{QuaDRiGa2}, the path gain of the generated channel trajectories is normalized.

\subsection{Measured Channel Data} \label{sec:meas_data}

\begin{figure}
    \centering
    \includegraphics[width=0.99\columnwidth]{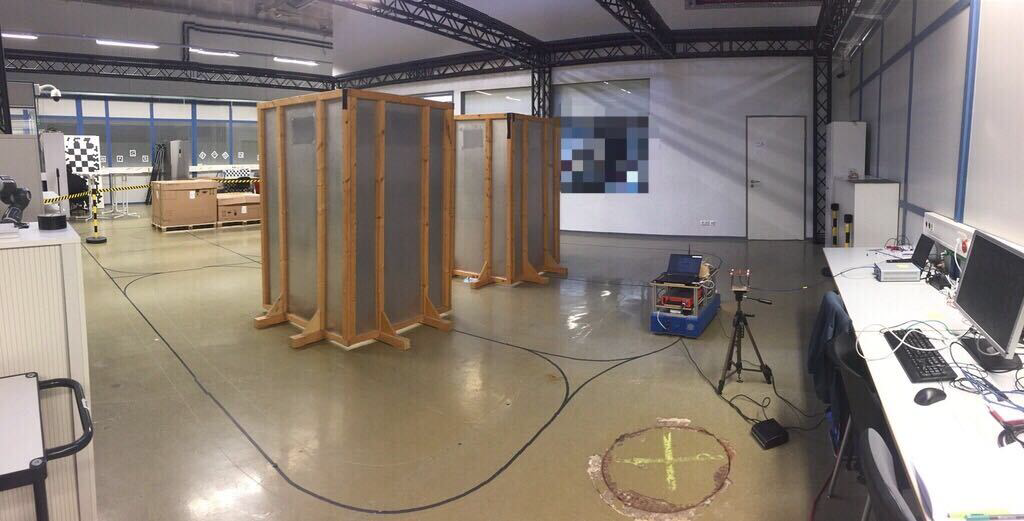}
    \caption{Industrial hall---Measurement site \cite{BuScHoFe21}. In this scenario, the blue AGV with attached measurement hardware moves around both obstacles.}
    \label{fig:tud_meas_area}
\end{figure}

Since synthetic channel data generally does not fully characterize the features of real-world environments, we utilize channel data measured in an industrial hall to assess the performance of the proposed channel predictor under real-world conditions.
A detailed description of the measurement campaign can be found in \cite{BuScHoFe21}. 
The dataset and supplementary material are available in \cite{BuSc23}.
The measurement area is modified to consider changing environments by placing obstacles at certain positions and capturing ten different measurement scenarios.
In \Cref{fig:tud_meas_area}, we depict exemplarily a measurement scenario where the \ac{AGV} moves around obstacles.
To facilitate generalization to different sub-carriers, which ranged from $\SI{3.7}{\giga\hertz}$ to $\SI{3.8}{\giga\hertz}$, and measurement scenarios, the training set and the test set consist of trajectories stemming from all available sub-carriers and measurement scenarios.
An \ac{AGV} moving along predefined trajectories with a fixed velocity of $v=\SI{1}{\meter/\second}$ mimics the moving \ac{MT}, whereby the symbol duration is $T_\mathrm{S}=\SI{1}{ms}$.
The issue outlined in \cite{BuJaTrScFe21}, concerning a fractional sampling time offset due to lack of synchronization between the transmitter and receiver, leads to an extra, subcarrier-specific phase drift over time in the channel frequency response.
With continuously measured data available, this phase drift can be tracked and corrected using linear regression in a post-processing step.

\section{Channel Prediction via GMMs}

The stochastic nature of all channel trajectories in the whole coverage area of the \ac{BS} is described by a \ac{PDF} \( f_{\mbh} \).
Every channel trajectory \( \mbh \) of any moving \ac{MT} within the environment is a realization of a random variable with \ac{PDF} \( f_{\mbh} \).
Motivated by the universal approximation property of \acp{GMM}~\cite{NgNgChMc20}, we make use of a \ac{GMM} to approximate the \ac{PDF} \( f_{\mbh} \) in an offline phase. 
Thereby, we learn the joint channel distribution of the observation and the prediction interval via the training dataset $\mathcal{H}$ defined in \eqref{eq:H_dataset} and, thus, are able to capture the dependencies between the observations and the \ac{CSI} values that should be predicted later in the inference/online phase. 

\subsection{Capturing the Environment -- Offline}

A \ac{GMM} with $K$ components is a \ac{PDF} of the following form: 
\begin{equation}\label{eq:gmm_of_h}
    f^{(K)}_{\mbh}(\mbh) = \sum\nolimits_{k=1}^K \pi_k \calN_{\C}(\mbh; \mbmu_k, \mbC_k).
\end{equation}
The \ac{GMM} components are described by the mixing coefficients $\pi_k$, means $\mbmu_k$, and covariances $\mbC_k$, where maximum likelihood estimates of these parameters can be obtained with an \ac{EM} algorithm given the training dataset \(\mathcal{H} \) in \eqref{eq:H_dataset}, cf.~\cite[Subsec.~9.2.2]{bookBi06}.

Based on the \ac{GMM}, the posterior probability (also called responsibility) that a particular channel trajectory stems from component \( k \) is given by~\cite[Sec.~9.2]{bookBi06}
\begin{equation}\label{eq:responsibilities_h}
    p(k \mid \mbh) = \frac{\pi_k \calN_{\C}(\mbh; \mbmu_k, \mbC_k)}{\sum_{i=1}^K \pi_i \calN_{\C}(\mbh; \mbmu_i, \mbC_i) }.
\end{equation}

The joint Gaussianity of each \ac{GMM} component [see \eqref{eq:gmm_of_h}] together with the \ac{AWGN} allows to compute the \ac{GMM} of the observations via [see \eqref{eq:signal_model} and \eqref{eq:sel_matrix}]
\begin{equation}\label{eq:gmm_y}
    f_{\mby}^{(K)}(\mby) = \sum\nolimits_{k=1}^K \pi_k \calN_{\C}(\mby; \mbS^\tp \meanhk, \mbS^\tp \covhk \mbS + \mbSigma).
\end{equation}
Accordingly, the responsibilities given noisy observations are computed as
\begin{equation}\label{eq:responsibilities}
    p(k \mid \mby) = \frac{\pi_k \calN_{\C}(\mby; \mbS^\tp \meanhk, \mbS^\tp \covhk \mbS + \mbSigma)}{\sum_{i=1}^K \pi_i \calN_{\C}(\mby; \mbS^\tp \meanhi, \mbS^\tp \covhi \mbS + \mbSigma) }.
\end{equation}

\subsection{Channel Prediction -- Online}

In the online phase, we aim to predict a single channel coefficient $h_\ell = h[M_\mathrm{o}-1+\ell]$, which lies $\ell$ steps in the future of the last observed channel coefficient $h[M_\mathrm{o}-1]$, with a given noisy observation $\mby$ from \eqref{eq:signal_model}. 
In this work, we try to approximate the \ac{CME}, given by
\begin{equation}
    \expec\left[ h_\ell \mid \mby\right]
    \label{eq:true_CME}
\end{equation}
which is the \ac{MSE}-optimal predictor \cite[Sec. 11.4]{Kay}.
Since we have learned a \ac{GMM} $f^{(K)}_{\mbh}$ for the joint distribution $f_{\mbh}$, we can compute the \ac{CME} over this approximate \ac{GMM} distribution, denoted by 
\begin{equation}
    \hat{h}_{\text{GMM}} = \expec^{(K)}\left[ h_\ell \mid \mby\right].
    \label{eq:approx_CME}
\end{equation}
Conveniently, this \ac{GMM}-based \ac{CME} can be computed in closed-form by utilizing the law of total expectation such that
\begin{align}
    \hat{h}_{\text{GMM}} &= \expec\left[ \expec^{(K)}[h_\ell \mid \mby, k ] \mid \mby\right]   \\
    &= \sum_{k=1}^{K} p(k \mid \mby) \expec^{(K)}\left[h_\ell \mid \mby, k \right] \label{eq:gmm_predictor}
\end{align}
with the responsibilities of the observations from \eqref{eq:responsibilities}.
Due to the Gaussianity of each \ac{GMM} component, the channel and observation become jointly Gaussian when conditioned on a \ac{GMM} component. Thus, the conditional expectation in \eqref{eq:gmm_predictor} 
is computed by \ac{LMMSE} predictions of the form
\begin{equation} \label{eq:gmm_lmmse_filters}
    \begin{aligned}
        \expec^{(K)}[&h_\ell \mid \mby, k ] = \mbc_{\mby,h_\ell|k}^\herm \mbC_{\mby|k}^{-1} (\mby\!-\! {\mbmu}_{\mby|k}) \!+ \mu_{h_{\ell}|k}\\
        &=\mathbf{e}_\ell^\tp\mbC_{k}\mbS(\mbS^\tp \mbC_{k} \mbS\!+\! \mbSigma)^{-1} (\mby\!-\! \mbS^\tp{\mbmu}_{k}) \!+ \mathbf{e}_\ell^\tp {\mbmu}_{k}
    \end{aligned}
\end{equation}
where $\mathbf{e}_\ell \in \{0,1\}^{M_\mathrm{o}+N_\mathrm{p}}$, which contains a one at the $\ell$-th entry and is zero elsewhere, cuts out the channel coefficient of interest since $h_\ell = \mathbf{e}_\ell^\tp\mbh$.
For a detailed analysis of the asymptotic behavior of the approximate \ac{CME} via the \ac{GMM} from \eqref{eq:approx_CME} and the true \ac{CME} from \eqref{eq:true_CME}, we refer the interested reader to \cite{KoFeTuUt21J}.
A considerable advantage of the \ac{GMM}-based prediction approach is that it does not require retraining for different \ac{SNR} levels since the \ac{GMM} of the observations in \eqref{eq:gmm_y} can be adapted depending on the noise statistics.

\subsection{Complexity Analysis}

The prediction filters $\mathbf{e}_\ell^\tp\mbC_{k}\mbS(\mbS^\tp \mbC_{k} \mbS\!+\! \mbSigma)^{-1}$ in \eqref{eq:gmm_lmmse_filters} can be precomputed offline for a given \ac{SNR} level since the \ac{GMM} parameters do not change after the fitting process.
Thus, evaluating the conditional expectation in \eqref{eq:gmm_lmmse_filters} has a complexity of only \( \mc{O}(M_\mathrm{o} + \ell) \).
To obtain a predicted channel coefficient, the computation of the responsibilities of the observations is further needed, which requires evaluating Gaussian densities, cf. \eqref{eq:responsibilities}.
Due to the fixed \ac{GMM} parameters, the online evaluation of the responsibilities is dominated by matrix-vector multiplications with a complexity of \( \mc{O}(M_\mathrm{o}^2) \).
Overall, evaluating \eqref{eq:gmm_predictor} has a total complexity of \( \calO(K M_\mathrm{o}^2) \).
Note that parallelization with respect to the number of components $K$ is possible.

\subsection{Structural Constraints} \label{sec:struct_gmm}

Due to model-based insights, the number of \ac{GMM} parameters can be reduced by constraining the \ac{GMM} covariance matrices.
Exploiting model-based insights has already been advantageous for applications such as channel estimation in \cite{FeJoHuKoTuUt22} and precoding in~\cite{TuFeUt23}.
In general, structural constraints reduce the number of parameters that need to be learned, lower the offline training complexity, and reduce the number of required training samples.
In this work, we will constrain the \ac{GMM} covariances to be Toeplitz, which can be expressed as
\begin{equation} \label{eq:struct_toep}
    \covhk = \mbQ^\herm \diag(\mbc_k) \mbQ
\end{equation}
where $\mbQ$ contains the first $M_\mathrm{o}+N_\mathrm{p}$ columns of a $2(M_\mathrm{o}+N_\mathrm{p})\times 2(M_\mathrm{o}+N_\mathrm{p})$ \ac{DFT} matrix, and $\mbc_k \in \R_{+}^{2(M_\mathrm{o}+N_\mathrm{p})}$ \cite{TuUt20, St86}.
Thus, the structural constraints allow storing only the vectors $\mbc_k$, $k=1,\dots,K$, of a \ac{GMM}, drastically reducing the memory requirement and the number of parameters to be learned without affecting the online prediction complexity compared to a \ac{GMM} with full covariance matrices.

\section{Baseline Channel Predictors}

We consider the following baseline channel predictors.
A predicted channel coefficient can be obtained, for example, with the \ac{LMMSE} predictor \cite{Kay}
\begin{equation} \label{eq:lmmse_filter}
    \hat{h}_{\text{LMMSE}}= \mathbf{e}_\ell^\tp\mbC_{s}\mbS(\mbS^\tp \mbC_{s} \mbS\!+\! \mbSigma)^{-1} \mby
\end{equation}
which utilizes the sample covariance matrix $\mbC_s = \frac{1}{J} \sum_{j=1}^J \mbh_j \mbh_j^\herm$ using the same channel trajectories which are used for fitting the \ac{GMM}, i.e., the training dataset \(\mathcal{H} \) [see \eqref{eq:H_dataset}] is used.

Alternatively, the \ac{LMMSE} predictor based on assuming that the spectrum of the signal follows Jakes' spectrum, which describes the temporal evolution of the channel in the one-ring channel model, cf. \cite{Zemen, Goldsmith, Kay}, is computed as
\begin{equation} \label{eq:lmmse_jakes_filter}
    \hat{h}_{\text{LMMSE Jakes}}= \mathbf{e}_\ell^\tp\mbC_{\text{Jakes}}\mbS(\mbS^\tp \mbC_{\text{Jakes}} \mbS\!+\! \mbSigma)^{-1} \mby.
\end{equation}
The covariance matrix $\mbC_{\text{Jakes}}$ is a Toeplitz-structured real-valued matrix, where its top row is given by the zeroth order Bessel function $ J_0(2\pi m T_\mathrm{S} f_\mathrm{c} {v}/{\mathrm{c}})$, cf.~\cite{Zemen, Goldsmith, Jakes}.
Note that the exact knowledge of the velocity $v$ is a prerequisite for the \ac{LMMSE} Jakes predictor.
We further consider the LMMSE Jakes predictor, where we assume either a $10\%$ or $20\%$ velocity estimation error since, in the online phase, a \ac{MT}'s velocity must be inferred from the noisy observations $\mby$.

Lastly, we compare to the \ac{NN}-based channel predictor from~\cite{TuUt20}, which infers a prediction filter from transformed noisy observations $\hat{\mbc} = {\frac{1}{\sigma^{2}}} |\mbQ_{\text{NN}}\mby|^2$ [$\mbQ_{\text{NN}}$ is similarly defined as in \eqref{eq:struct_toep}] at the input of the \ac{NN} and applies it to the observations to compute a predicted coefficient:
\begin{equation}
    \hat{h}_{\text{NN}} = \hat{\mbw}^\tp(\hat{\mbc}) \mby.
\end{equation}
A drawback of the \ac{NN}-based prediction approach is that it requires training for each \ac{SNR} level.
Details about the network architecture and the hyper-parameters can be found in \cite{TuUt20}.

\begin{figure}[t]
    \centering
        \begin{tikzpicture}
\begin{axis}[
legend style={font=\scriptsize},
  width=0.99\columnwidth, 
  height=0.8\columnwidth,  
  grid = both,
  ymode=log,log basis y=10,
  xmin = -10,   
  xmax = 30,  
  ymin = 5e-3,
  ymax = 1,
  /pgfplots/xtick = {-10,-5,...,30},  
  axis background/.style = {fill=white},
  legend style={at={(0.0,0.0)},anchor=south west},
  ylabel = {MSE},
  xlabel = {SNR [$\SI{}{\deci\bel}$]},
  tick align = inside,]
\addplot[gmm] table [x=snr_dB, y=GMM_full, col sep=comma] {Figures/graph_81.csv};
\addlegendentry{GMM};
\addplot[gmmtoep] table [x=snr_dB, y=GMM_toep, col sep=comma] {Figures/graph_81.csv};
\addlegendentry{Toeplitz GMM};
\addplot[lmmse] table [x=snr_dB, y=samp_cov, col sep=comma] {Figures/graph_81.csv};
\addlegendentry{LMMSE};
\addplot[cnn] table [x=snr_dB, y=NN_toep, col sep=comma] {Figures/graph_81.csv};
\addlegendentry{NN};
\addplot[jakes] table [x=snr_dB, y=Jakes, col sep=comma] {Figures/graph_81.csv};
\addlegendentry{LMMSE Jakes Perfect};
\addplot[jakes10] table [x=snr_dB, y=Jakes_10_dev, col sep=comma] {Figures/graph_81.csv};
\addlegendentry{LMMSE Jakes ($\pm10\%$)};
\addplot[jakes20] table [x=snr_dB, y=Jakes_20_dev, col sep=comma] {Figures/graph_81.csv};
\addlegendentry{LMMSE Jakes ($\pm20\%$)};

\end{axis}
\end{tikzpicture}
    \vspace{-0.5cm}
    \caption{Channel prediction error over the \ac{SNR} for a system with $M_\mathrm{o}=19$, $\ell=1$, and $K=128$ (synthetic channel data).}
    \label{fig:overSNR_M19_ell1}
\end{figure}
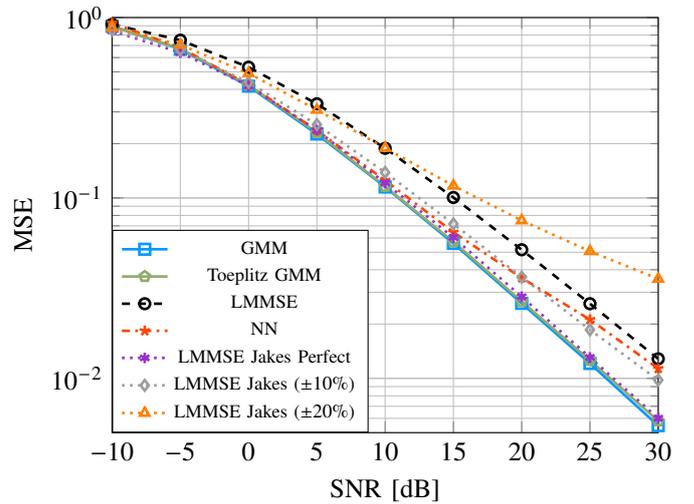

\section{Experiments and Results}
\label{sec:sim_results}

In our simulations, the \ac{SNR} is defined as \( \frac{1}{\sigma^2} \), since we normalized the data such that $\expec[\|\mbh\|^2] = M_\mathrm{o}+N_\mathrm{p}$.
To assess the prediction performance, we utilize the \ac{MSE} \( \frac{1}{T} \sum_{{t}=1}^{T} | h_t[m] - \hat{h}_t[m] |^2\) as performance measure and use $T=10{,}000$ test samples.
We utilize $J=150{,}000$ training samples for fitting the \ac{GMM} and for training the \acp{NN}.

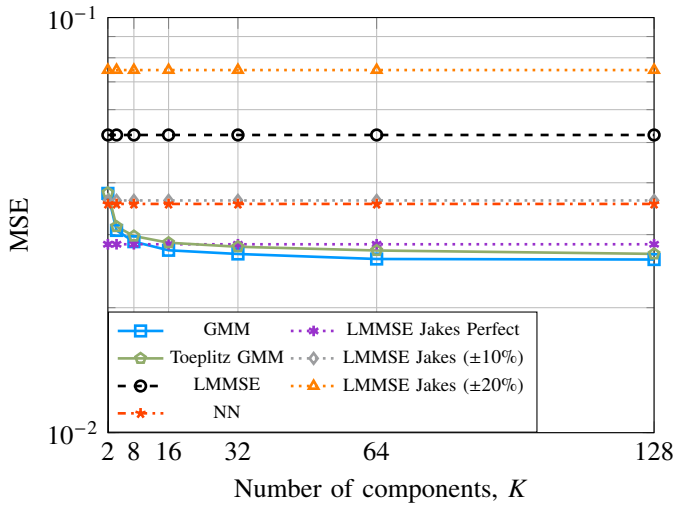
\begin{figure}[t]
    \centering
        \begin{tikzpicture}
\begin{axis}[
legend style={font=\scriptsize},
  width=0.99\columnwidth, 
  height=0.8\columnwidth,
  grid = both,
  ymode=log,log basis y=10,
  xmin = 2,   
  xmax = 128,  
  ymin = 1e-2,   
  ymax = 1e-1, 
  xtick = {2,8,16,32,64,128},
  axis background/.style = {fill=white},
  legend style={at={(0.0,0.0)},anchor=south west, legend columns=2},
  ylabel = {MSE},
  xlabel = {Number of components, $K$},
  tick align = inside,]
\addplot[gmm] table [x=comp, y=GMM_full, col sep=comma] {Figures/graph_88.csv};
\addlegendentry{GMM};
\addplot[jakes] table [x=comp, y=Jakes, col sep=comma] {Figures/graph_88.csv};
\addlegendentry{LMMSE Jakes Perfect};
\addplot[gmmtoep] table [x=comp, y=GMM_toep, col sep=comma] {Figures/graph_88.csv};
\addlegendentry{Toeplitz GMM};
\addplot[jakes10] table [x=comp, y=Jakes_10_dev, col sep=comma] {Figures/graph_88.csv};
\addlegendentry{LMMSE Jakes ($\pm10\%$)};
\addplot[lmmse] table [x=comp, y=samp_cov, col sep=comma] {Figures/graph_88.csv};
\addlegendentry{LMMSE};
\addplot[jakes20] table [x=comp, y=Jakes_20_dev, col sep=comma] {Figures/graph_88.csv};
\addlegendentry{LMMSE Jakes ($\pm20\%$)};
\addplot[cnn] table [x=comp, y=NN_toep, col sep=comma] {Figures/graph_88.csv};
\addlegendentry{NN};
\end{axis}
\end{tikzpicture}
    \vspace{-0.5cm}
    \caption{Channel prediction error over the number of components $K$, for a system with $M_\mathrm{o}=19$, $\ell=1$, and \ac{SNR} $=\SI{20}{\deci\bel}$ (synthetic channel data).}
    \label{fig:overComponents_M19_ell1_comps}
\end{figure}

\subsection{Experiments with Synthetic Channel Data}

Firstly, we present simulation results using the synthetic channel data described in Subsection \ref{sec:syn_data}.
In \Cref{fig:overSNR_M19_ell1}, we depict the prediction error over the \ac{SNR}, for an observation length of $M_\mathrm{o}=19$, and we predict one step into the future, i.e., $\ell=1$.
We set the number of \ac{GMM} components to $K=128$.
We can observe that the \ac{GMM} performs best, followed by the \ac{GMM} with the Toeplitz structure enforced on the covariance matrices (denoted by ``Toeplitz GMM'').
The ``LMMSE Jakes Perfect'' predictor performs almost equally well but, in contrast to the \ac{GMM}-based predictor, requires the exact knowledge of the \ac{MT}'s velocity. 
We can observe a severe performance degradation by artificially introducing a velocity estimation error, cf. ``LMMSE Jakes ($\pm10\%$)'' and ``LMMSE Jakes ($\pm20\%$)''.
The \ac{NN} approach (denoted by ``NN'') performs worse than the \ac{GMM} and achieves a similar performance as ``LMMSE Jakes ($\pm10\%$)''.
The \ac{NN} approach also does not require the knowledge of the velocity as the \ac{GMM}-based predictor but requires separate training for each \ac{SNR} level.
In principle, a \ac{NN} for the whole \ac{SNR} range can be learned, but a \ac{NN} trained for a large range of \ac{SNR} levels generally exhibits a worse performance, cf. e.g., \cite{MaGü21}.

In \Cref{fig:overComponents_M19_ell1_comps}, we simulate the same setup but fix the \ac{SNR} to $\SI{20}{\deci\bel}$ and vary the number of components $K$ of the \ac{GMM}-based predictors.
Accordingly, all baselines remain constant since they do not depend on $K$.
With an increasing number of components, the prediction performance steadily increases and exhibits a saturation for $K>16$.
Both of the \ac{GMM}-based predictors outperform most baselines with only a few components.
We can observe that the Toeplitz constraint comes at the cost of slightly degraded performance and requires approximately $32$ components to outperform ``LMMSE Jakes Perfect'' as opposed to the \ac{GMM} with full covariances, which only requires $16$ components.

\subsection{Experiments with Measured Channel Data}

\begin{figure}[t]
    \centering
        \begin{tikzpicture}
\begin{axis}[
legend style={font=\scriptsize},
  width=0.99\columnwidth, 
  height=0.8\columnwidth,  
  grid = both,
  ymode=log,log basis y=10,
  xmin = -10,   
  xmax = 30,  
  ymin = 1e-3,
  ymax = 7e-1,
  /pgfplots/xtick = {-10,-5,...,30},  
  axis background/.style = {fill=white},
  legend style={at={(0.0,0.0)},anchor=south west},
  ylabel = {MSE},
  xlabel = {SNR [$\SI{}{\deci\bel}$]},
  tick align = inside,]
\addplot[gmm] table [x=snr_dB, y=GMM_full, col sep=comma] {Figures/graph_81_TUD.csv};
\addlegendentry{GMM};
\addplot[gmmtoep] table [x=snr_dB, y=GMM_toep, col sep=comma] {Figures/graph_81_TUD.csv};
\addlegendentry{Toeplitz GMM};
\addplot[lmmse] table [x=snr_dB, y=samp_cov, col sep=comma] {Figures/graph_81_TUD.csv};
\addlegendentry{LMMSE};
\addplot[cnn] table [x=snr_dB, y=NN_toep_P_2, col sep=comma] {Figures/graph_81_TUD.csv};
\addlegendentry{NN};
\addplot[jakes] table [x=snr_dB, y=Jakes, col sep=comma] {Figures/graph_81_TUD.csv};
\addlegendentry{LMMSE Jakes Perfect};
\addplot[jakes10] table [x=snr_dB, y=Jakes_10_dev, col sep=comma] {Figures/graph_81_TUD.csv};
\addlegendentry{LMMSE Jakes ($\pm10\%$)};
\addplot[jakes20] table [x=snr_dB, y=Jakes_20_dev, col sep=comma] {Figures/graph_81_TUD.csv};
\addlegendentry{LMMSE Jakes ($\pm20\%$)};

\end{axis}
\end{tikzpicture}
    \vspace{-0.5cm}
    \caption{Channel prediction error over the \ac{SNR} for a system with $M_\mathrm{o}=19$, $\ell=1$, and $K=128$ (measured channel data).}
    \label{fig:overSNR_M19_ell1_meas}
\end{figure}
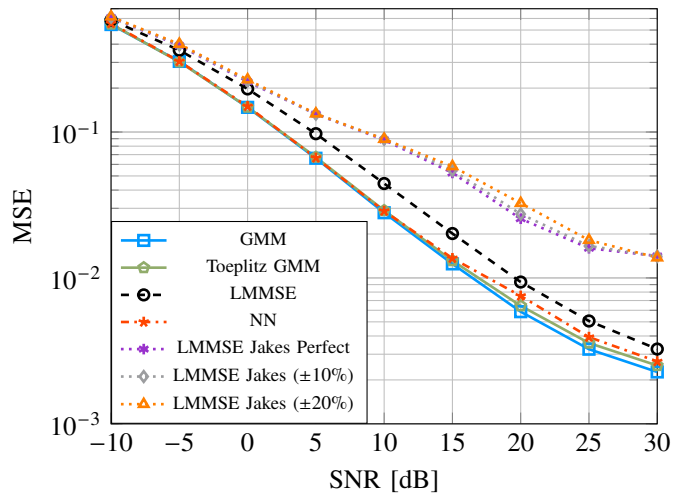

In the remainder, we focus our numerical evaluation on the measured channel data described in Subsection \ref{sec:meas_data}.
In \Cref{fig:overSNR_M19_ell1_meas}, we depict the prediction error over the \ac{SNR} for an observation length of $M_\mathrm{o}=19$, and we predict one step into the future, i.e., $\ell=1$.
We set the number of \ac{GMM} components to $K=128$.
In alignment with the results for the synthetic data, also in the case of measured data, the \ac{GMM} performs best, followed by the \ac{GMM} with the Toeplitz structure enforced on the covariance matrices (``Toeplitz GMM'').
In contrast to the synthetic channel data simulations, we can observe a huge performance degradation of the ``LMMSE Jakes Perfect'' predictor, due to the one-ring model assumption of Jakes, which is not fulfilled in the indoor factory hall measurement environment, especially, in the presence of a \ac{LOS} condition.
The artificially introduced velocity estimation errors, cf. ``LMMSE Jakes ($\pm10\%$)'' and ``LMMSE Jakes ($\pm20\%$)'' lead to further performance degradations in this case.
Furthermore, the \ac{NN} approach with \ac{SNR} level-specific training (``NN'') performs slightly worse than the \ac{GMM}.

\begin{figure}[t]
    \centering
        \begin{tikzpicture}
\begin{axis}[
legend style={font=\scriptsize},
  width=0.99\columnwidth, 
  height=0.8\columnwidth,  
  grid = both,
  ymode=log,log basis y=10,
  xmin = -10,   
  xmax = 30,  
  ymin = 6e-3,
  ymax = 7e-1,
  /pgfplots/xtick = {-10,-5,...,30},  
  axis background/.style = {fill=white},
  legend style={at={(0.0,0.0)},anchor=south west},
  ylabel = {MSE},
  xlabel = {SNR [$\SI{}{\deci\bel}$]},
  tick align = inside,]
\addplot[gmm] table [x=snr_dB, y=GMM_full, col sep=comma] {Figures/graph_80_TUD.csv};
\addlegendentry{GMM};
\addplot[gmmtoep] table [x=snr_dB, y=GMM_toep, col sep=comma] {Figures/graph_80_TUD.csv};
\addlegendentry{Toeplitz GMM};
\addplot[lmmse] table [x=snr_dB, y=samp_cov, col sep=comma] {Figures/graph_80_TUD.csv};
\addlegendentry{LMMSE};
\addplot[cnn] table [x=snr_dB, y=NN_toep_P_2, col sep=comma] {Figures/graph_80_TUD.csv};
\addlegendentry{NN};
\addplot[jakes] table [x=snr_dB, y=Jakes, col sep=comma] {Figures/graph_80_TUD.csv};
\addlegendentry{LMMSE Jakes Perfect};
\addplot[jakes10] table [x=snr_dB, y=Jakes_10_dev, col sep=comma] {Figures/graph_80_TUD.csv};
\addlegendentry{LMMSE Jakes ($\pm10\%$)};
\addplot[jakes20] table [x=snr_dB, y=Jakes_20_dev, col sep=comma] {Figures/graph_80_TUD.csv};
\addlegendentry{LMMSE Jakes ($\pm20\%$)};

\end{axis}
\end{tikzpicture}
    \vspace{-0.5cm}
    \caption{Channel prediction error over the \ac{SNR} for a system with $M_\mathrm{o}=16$, $\ell=4$, and $K=128$ (measured channel data).}
    \label{fig:overSNR_M16_ell4_meas}
\end{figure}
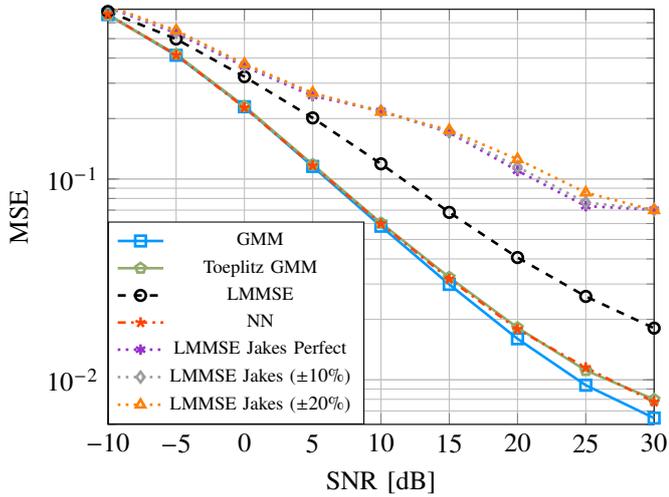

In \Cref{fig:overSNR_M16_ell4_meas}, we depict the prediction error over the \ac{SNR}, for an observation length of $M_\mathrm{o}=16$, and we predict four steps into the future, i.e., $\ell=4$, and keep $K=128$.
We can observe that compared to the simulation setting in \Cref{fig:overSNR_M19_ell1_meas}, all of the approaches perform worse since the prediction task is harder with fewer observations ($M_\mathrm{o}=16$ instead of $19$) and with a larger prediction step ($\ell=4$ instead of $1$).
The ordering of the performances of the prediction approaches remains the same, but the performance gap of the \ac{GMM}-based predictor compared to the ``LMMSE'' and ``LMMSE Jakes Perfect'' increased.

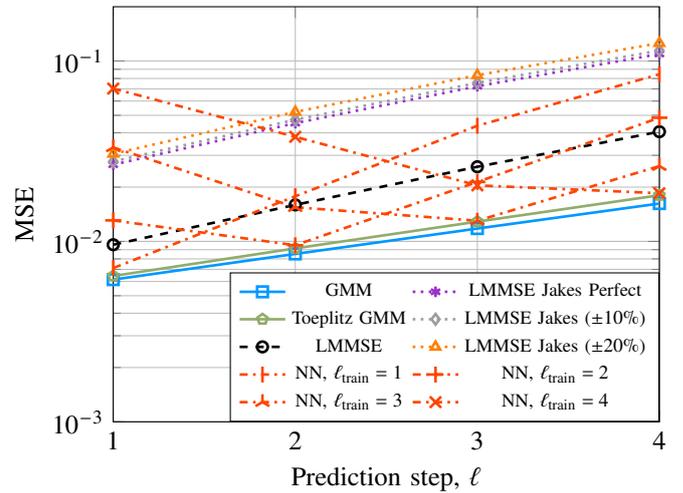
\begin{figure}[t]
    \centering
        \begin{tikzpicture}
\begin{axis}[
legend style={font=\scriptsize},
  width=0.99\columnwidth, 
  height=0.8\columnwidth,
  grid = both,
  ymode=log,log basis y=10,
  xmin = 1,   
  xmax = 4,  
  ymin = 1e-3,   
  ymax = 2e-1, 
  xtick = {1,2,3,4},
  axis background/.style = {fill=white},
  legend style={at={(1.0,0.0)},anchor=south east, legend columns=2},
  ylabel = {MSE},
  xlabel = {Prediction step, $\ell$},
  tick align = inside,]
\addplot[gmm] table [x=l, y=GMM_full, col sep=comma] {Figures/graph_91_TUD_SNR_20.csv};
\addlegendentry{GMM};
\addplot[jakes] table [x=l, y=Jakes, col sep=comma] {Figures/graph_84_TUD_SNR_20.csv};
\addlegendentry{LMMSE Jakes Perfect};
\addplot[gmmtoep] table [x=l, y=GMM_toep, col sep=comma] {Figures/graph_91_TUD_SNR_20.csv};
\addlegendentry{Toeplitz GMM};
\addplot[jakes10] table [x=l, y=Jakes_10_dev, col sep=comma] {Figures/graph_84_TUD_SNR_20.csv};
\addlegendentry{LMMSE Jakes ($\pm10\%$)};
\addplot[lmmse] table [x=l, y=samp_cov, col sep=comma] {Figures/graph_84_TUD_SNR_20.csv};
\addlegendentry{LMMSE};
\addplot[jakes20] table [x=l, y=Jakes_20_dev, col sep=comma] {Figures/graph_84_TUD_SNR_20.csv};
\addlegendentry{LMMSE Jakes ($\pm20\%$)};
\addplot[cnn, mark=|, mark size=3.0, opacity=1.0] table [x=l, y=NN_toep_step_1_P_1, col sep=comma] {Figures/graph_91_TUD_SNR_20.csv};
\addlegendentry{NN, $\ell_{\text{train}} = 1$};
\addplot[cnn, mark=+, mark size=3.0, opacity=1.0] table [x=l, y=NN_toep_step_2_P_2, col sep=comma] {Figures/graph_91_TUD_SNR_20.csv};
\addlegendentry{NN, $\ell_{\text{train}} = 2$};
\addplot[cnn, mark=Mercedes star, mark size=3.0, opacity=1.0] table [x=l, y=NN_toep_step_3_P_2, col sep=comma] {Figures/graph_91_TUD_SNR_20.csv};
\addlegendentry{NN, $\ell_{\text{train}} = 3$};
\addplot[cnn, mark=x, mark size=3.0, opacity=1.0] table [x=l, y=NN_toep_step_4_P_2, col sep=comma] {Figures/graph_91_TUD_SNR_20.csv};
\addlegendentry{NN, $\ell_{\text{train}} = 4$};
\end{axis}
\end{tikzpicture}
    \vspace{-0.5cm}
    \caption{Channel prediction error over the prediction step $\ell$, for a system with $M_\mathrm{o}=16$, $K=128$, and \ac{SNR} $=\SI{20}{\deci\bel}$ (measured channel data).}
    \label{fig:overell_M16_SNR20}
\end{figure}

Lastly, in \Cref{fig:overell_M16_SNR20}, we evaluate a setup with $M_\mathrm{o}=16$, $K=128$, fix the \ac{SNR} at $\SI{20}{\deci\bel}$, and vary the prediction step $\ell$.
As explained above, the \ac{GMM}-based prediction approach is trained once for the overall trajectory of length $20$ and is customized to the different prediction steps $\ell \in \{1,2,3,4\}$.
It yields the best performance, followed by the \ac{GMM} with the Toeplitz structure enforced on the covariance matrices (``Toeplitz GMM'').
With ``NN, $\ell_{\text{train}} \in \{1,2,3,4\}$'', we denote the \ac{NN} approaches trained for fixed prediction steps $\ell_{\text{train}}$, where the obtained predicted coefficient in the online phase is taken as representative for the prediction step $\ell$ of interest.
We can see that each \ac{NN} trained for a specific prediction step $\ell_{\text{train}}$ performs best if the prediction step of interest $\ell$ is equal to $\ell_{\text{train}}$, thus, highlighting the need for a \ac{NN} specifically trained for a particular prediction step in addition to the \ac{SNR} level.

\section{Conclusion and Future Work}

We proposed a wireless channel predictor based on \acp{GMM} and assessed its performance with synthetic and measured channel data.
Once the \ac{GMM} is trained for a predefined trajectory length in the offline phase, it can be customized for varying observations and prediction lengths in the online phase without requiring retraining for different \ac{SNR} levels.
The \ac{GMM}-based channel predictor even allows for multiple predictions simultaneously by adapting the prediction filters in \eqref{eq:gmm_lmmse_filters} accordingly.
Future work aims to explore an extension to systems with multiple antennas and systems involving coarse quantization, cf., e.g., \cite{FeTuBoUt23, TuKoUt21}.
Moreover, other generative modeling-based techniques, such as variational autoencoders, could be utilized as generative priors for channel prediction similar to the channel estimation case as in \cite{BoBaRiUt23}.

\bibliographystyle{IEEEtran}
\bibliography{IEEEabrv,biblio}
\balance

\end{document}